\documentclass[a4paper,12pt]{article}

\usepackage{amsmath}
\usepackage{amssymb}
\usepackage{amsthm}
\usepackage{bbm}
\usepackage{xcolor}
\usepackage{graphicx}

\usepackage[english]{babel}
\newtheorem{rem}{Remark}

\title{Entanglement and non-locality in quantum protocols with identical particles}

\author{F. Benatti$^{a,b}$, R. Floreanini$^b$, U. Marzolino$^b$ \\
\\
\small ${}^a$Department of Physics, University of Trieste, 34151 Trieste, Italy \\
\small ${}^b$Istituto Nazionale di Fisica Nucleare (INFN), Sezione di Trieste, 34151 Trieste, Italy}
\date{}

\begin{document}

\maketitle

\begin{abstract}
We study the role of entanglement and non-locality in quantum protocols that make use of systems of identical particles. Unlike in the case of distinguishable particles, the notions of entanglement and non-locality for systems whose constituents cannot be distinguished and singly addressed are still debated. We clarify why the only approach that avoids incongruities and paradoxes is the one based on the second quantization formalism whereby it is the entanglement of the modes that can be populated by the particles what really matters and not the particles themselves. Indeed, by means of a metrological and of a teleportation protocol, we show that inconsistencies arise in formulations that force entanglement and non-locality to be properties of the identical particles rather than of the modes they can occupy. The reason resides in the fact that orthogonal modes can always be addressed while identical particles can not.
\end{abstract}

\section{Introduction}
\label{intro}

Entanglement is the strongest among quantum correlations, rooted in the structurally non-local behaviour of quantum mechanics.
It is a fundamental resource in most quantum information protocols and processes, as shown in actual applications
using different physical settings, {\it e.g.} in quantum optics and atomic and molecular physics \cite{Amico2008,Tichy2011,Pan2012,Fabre2020,Polino2020,Genovese2021}. 

Although well established for systems of distinguishable particles \cite{Werner1989,Horodecki2009}  the notions of non-locality and entanglement are challenging in systems of indistinguishable particles, as in this case particles cannot be individually addressed and measured. 
On the other hand, many-body systems, made of identical constituents, are central to condensed matter physics
and therefore of utmost importance in actual applications. Our aim is to provide instances in which the implementation of
quantum protocols with systems of identical particles results particularly effective, possibly stimulating a discussion on
their actual experimental realization.

Many-body systems are more easily treated adopting the so-called ``second-quantization'' approach; within this framework, the notions of non-locality and entanglement valid for distinguishable particles 
can be easily generalized to hold in systems made of identical constituents, in a way applicable to all physical situations.
Although quantum enhanced performances
of technological tasks can be achieved by exploiting other quantum resources
than solely entanglement \cite{Braun2018,Chitambar2019}, in the following we focus on quantum protocols that require entanglement and
are implemented by means of systems consisting of many identical particles.
Specifically, we shall analyze two quantum protocols, one in quantum {\it metrology} and the other one in {\it teleportation}, 
and describe in detail how for identical particle systems non-locality and entanglement can be used to reach accuracies beyond those obtainable with classical methods.

Concerning the first protocol, a paradigmatic quantum metrological task is the estimation of interferometric phases 
\cite{Helstrom1976,Holevo1982,Braunstein1996,Paris2009,Braun2018} which has applications for the detection of gravitational waves \cite{Tse2019}, angular rotation \cite{Fink2019,Moan2020,Ding2020,Grace2020,Wu2020}, and temperature \cite{Stace2010}. As we shall see, the difference between distinguishable and indistinguishable particles is that the former require initial state entanglement to beat classical performances, while the latter achieve quantum enhancements with initial separable states as the same interferometers themselves can generate the necessary indistinguishable
particle entanglement \cite{Benatti2010,Benatti2011}. Thus particle indistinguishability mitigates the needed efforts in actual experimental implementations.

On the other hand, the second protocol, quantum teleportation, is known to play a prominent role in quantum computation \cite{Gottesman1999,Zhou2000,Childs2005,Gross2007,Chiu2013,Hillmich2020} and quantum communication \cite{Wilde,Khatri,Briegel1998,Dur1999}.
In this respect, a particular application of teleportation is the so-called {\it entanglement swapping}, that is the generation of quantum correlations between distant subsystems
using their entanglement with an auxiliary system:
as discussed below, it can be effectively implemented using
identical particle systems.

Let us point out that recent technological advances in quantum optics, \cite{Pan2012,Fabre2020,Polino2020,Li2020,Tschernig2020,GarciaEscartin2020,Genovese2021} as well as in ultracold atom physics \cite{Wurtz2009,Bakr2010,Sherson2010,Schlederer2020} make these protocols likely in the reach of present experimental abilities, thus paving the way to 
the adoption of many-body systems, made of indistinguishable constituents, as the standard framework for
the actual realization of quantum information tasks. Indeed, identical particle systems offer clear advantages 
in implementing quantum informational protocols with respect to the more standard realizations using distinguishable qubits.
First of all, using systems of singly addressable constituents one needs to keep them always well localized in order to preserve
their distinguishability, and this requires more resources. Moreover, as already mentioned before, in order to get accuracies beyond the classical limit with systems of
identical particles, there is no need of initially prepare them in an entangled state before feeding them
to an apparatus implementing a specific quantum task.

Although we focus here on the notion of entanglement for indistinguishable particle systems in the 
second quantized framework (the so called {\it mode entanglement})~\cite{Zanardi2001,Bose2002,Zanardi2002-1,Calsamiglia2002,Shi2003,Schuch2004-1,Narnhofer2004}, different approaches to identical particle entanglement
have been proposed in the literature (see \cite{Benatti2020} for a complete review). They have been introduced having in mind
specific physical models and applications to limited quantum tasks; in a way or the other, they fail to provide a comprehensive and at the same time fully consistent theory
of identical particle entanglement valid in all situations. Indeed, the different definitions of entanglement,
alternative to mode entanglement, may produce contradictory results {\it e.g.} in metrological applications,
where beating the classical limit to the reachable accuracy is apparently possible without any entanglement ether in the initial state or generated during the protocol.

On the other hand, mode entanglement \cite{Benatti2017,Benatti2020,Johann2020}, the approach adopted in the present investigation, has been proven to be free of such inconsistencies. As discussed in detail below, in this approach
the subsystems to be entangled are identified with modes in second quantization rather than (unaddressable) particles, a natural approach in quantum optics \cite{Pan2012,Fabre2020}, quantum field theory \cite{Emch,Narnhofer2002,Rangamani2017,Holland2018,Nishioka2018,Witten2018}, and quantum statistical mechanics \cite{Emch}.

In the second part of the work we examine different approaches to indistinguishable particle entanglement alternative to mode entanglement; using established results provided by the literature, we discuss in detail, through explicit examples, how these approaches fail to properly identify quantum resources as a consequence of their inability to consistently cope with the locality issue.

The structure of the paper is as follows.
In Section \ref{mode-ent-sec}, we introduce the theory of mode-entanglement 
its motivations, techniques and applications. In Section \ref{protocols-sec}, the two benchmark protocols, {\it i.e.} interferometric phase estimation and teleportation, are discussed within the framework of mode-entanglement. Then,  in Section \ref{disc-sec}, we compare the predictions relative to the two protocols provided by the mode-entanglement point of view with those of some alternative approaches. Conclusions are finally drawn in Section \ref{conclusions}, 
while more technical details are provided in the Appendices.

We do hope that the present investigation will reinforce the attention on the advantages of using many-body systems
for performing quantum informational tasks, especially in view of possible experimental realizations.

\section{Mode entanglement}
\label{mode-ent-sec}

The main reason behind the existence of different notions of identical particle entanglement and  their inconsistencies 
is the attempt of most of them to hinge on the particle picture in a context where particles cannot be individually addressed and measured.
Clearly, the focus on the particle aspect is prompted by entanglement theory as it has been developed for systems made of distinguishable particles. In order to appreciate the kind of difficulties that arise by pursuing too close an approach to the standard one, let us consider two Bosons prepared in two orthogonal states $\vert\chi\rangle$ and $\vert\phi\rangle$.  Because of their indistinguishability, we can only say that one Boson is in the state vector $\vert\chi\rangle$ and the other one in the state vector $\vert\phi\rangle$ orthogonal to $\vert\chi\rangle$, but we cannot attribute to any of them a specific particle label.
The corresponding  normalized state must indeed be symmetric,
\begin{equation}
\label{eq1}
\vert\Psi\rangle=\frac{\vert\chi\rangle\otimes|\phi\rangle\,+\,\vert\phi\rangle\otimes|\chi\rangle}{\sqrt{2}}\ ,
\end{equation}
so that neither $\vert\chi\rangle$ nor $\vert\phi\rangle$ can be specifically attributed as a state to a specific particle. 
In the case of two distinguishable particles, $\vert\Psi\rangle$ would be an entangled state; is it so also for two Bosons?

This question has a counterpart that involves operators. Indeed, bipartite entanglement is strictly related to the notion of locality of observables. In practice, given two distinguishable particles, labelled by $1$ and $2$, one deals with observables of particle $1$ of the form $X\otimes \mathbbm{1}$, belonging to the set $\mathcal{O}_1$, respectively of particle $2$ of the form $\mathbbm{1}\otimes Y$, in the corresponding set $\mathcal{O}_2$. These two sets of operators are subalgebras
of the algebra $\mathcal{O}=\mathcal{O}_1 \cup \mathcal{O}_2$ containing the observables of the two-particle system. In particular, two-particle observables of the following form, {\it i.e.} product of two single particle operators,
\begin{equation}
\label{eq5}
\big(X\otimes\mathbbm{1}\big)\big(\mathbbm{1}\otimes Y\big)=X\otimes Y\ ,
\end{equation}
are called \textit{local}; indeed all elements of $\mathcal{O}_1$ commute with all elements of $\mathcal{O}_2$ whence local measurements of operators from $\mathcal{O}_1$ and $\mathcal{O}_2$ do not influence each other. 
It turn out that entangled state vectors  of the two-particles are those $\vert\Psi\rangle$  that violate the factorization of correlation functions of local observables:
\begin{equation}
\label{eq6}
\langle\Psi\vert X\otimes Y\vert\Psi\rangle\,\neq\,\langle\Psi\vert X\otimes \mathbbm{1}\vert\Psi\rangle\,\langle\Psi\vert \mathbbm{1}
\otimes Y\vert\Psi\rangle\ ,
\end{equation}
for at least one commuting pair $X\in\mathcal{O}_1$ and $Y\in\mathcal{O}_2$. 
Notice that the equality in the previous expression would witness the statistical independence of the two observables with respect to the given bipartite state vector $\vert\Psi\rangle$.
One easily checks that the entanglement of a state as $\vert\Psi\rangle$ in~\eqref{eq1} is then witnessed by the local observable 
$P_\chi\otimes P_\phi$, where $P_\chi$, respectively $P_\phi$ project onto $\vert\chi\rangle$, respectively $\vert\phi\rangle$,
$$
\langle\Psi\vert P_\chi\otimes P_\phi\vert\Psi\rangle=\frac{1}{2}\neq \langle\Psi\vert P_\chi\otimes \mathbbm{1}\vert\Psi\rangle\,\langle\Psi\vert \mathbbm{1}\otimes P_\phi\vert\Psi\rangle=\frac{1}{4}\ .
$$
In a nutshell, distinguishable particle pure state entanglement can be understood as the ability of certain state vectors 
of making statistically dependent certain particle observables that are algebraically independent, namely commuting.\\
Unfortunately, tensor products as $X\otimes\mathbbm{1}$, $\mathbbm{1}\otimes Y$ and  $X\otimes Y$  cannot be sensible operators for identical particles.
Indeed, they attach an operator to a specific particle, $X$ to particle $1$ and $Y$ to 
particle $2$, making them distinguishable; in order to make those tensor products meaningful observables for indistinguishable particles, one needs symmetrize them yielding
\begin{equation}
\label{eq2a}
X_{\textnormal{sym}}\equiv X\otimes \mathbbm{1}\,+\,\mathbbm{1}\otimes X\ ,\quad Y_{\textnormal{sym}}\equiv Y\otimes \mathbbm{1}\,+\,\mathbbm{1}\otimes Y\ ,
\end{equation}
for the single-particle operators and
\begin{equation}
\label{eq2}
C=X\otimes Y\,+\,Y\otimes X\ ,
\end{equation}
for the two-particle ones. Only in this way one avoids the association of a specific particle label either to $X$ or $Y$.
However, unlike $X\otimes Y$, $C$ is non-local from the distinguishable particle point of view: 
does the same conclusion hold true for identical particles?\\
In other words, the question is whether symmetrization, or anti-symmetrization for Fermions, automatically generates entanglement and non-locality.

\begin{rem}
\label{loc:rem}
There is however a further and deeper issue connected with the notion of locality for identical particles when one tries to formulate it in terms of tensor products of single particle commuting operators. Although the single particle operators $X_{\textnormal{sym}}$ and $Y_{\textnormal{sym}}$ commute if and only if $[X,Y]=0$, their products yield operators of the form $C$ in~\eqref{eq2} if and only if $XY=0$. There are thus plenty of two-particle observables that one would deem local, but cannot be expressed as products of single particle observables. Such an observation is at the root of the fact that identical particle locality cannot be consistently formulated hinging upon the notion of particles and upon the ensuing description in terms of tensor products either of states or of operators~\cite{Benatti2020}. A satisfactory way out from this puzzle as well as the answers to the two questions raised above come as follows.
\end{rem}

Though identical particles can be addressed within the particle-based formalism of first quantization, that labels particle degrees of freedom as exemplified in~\eqref{eq1}, the most natural tools for describing the physics of many-body systems are those of second quantization.
In second quantization, one focusses upon modes, namely single particle state vectors $\psi$, and upon their creation $a^\dag_\psi$ and annihilation operators $a_\psi$, instead of upon particles themselves. 
Acting on the vacuum $\vert \textnormal{vac}\rangle$, such that $a_\psi\vert \textnormal{vac}\rangle=0$,  $a_\psi^\dag$ creates  a particle in the state $\vert\psi\rangle$, while its adjoint $a_\psi$ destroys a particle in the state $\psi$:
\begin{equation}
\label{eq2aa}
a^\dag_\psi\vert \textnormal{vac}\rangle=\vert\psi\rangle\ ,\qquad a_\psi\vert\psi\rangle =\vert \textnormal{vac}\rangle\ .
\end{equation}
These operators satisfy the canonical commutation relations for Bosons 
\begin{equation}
\label{eq3}
[a_\psi,a^\dag_\varphi]\equiv a_\psi\,a^\dag_\varphi\,-\,a^\dag_\varphi\,a_\psi=\langle\psi\vert\varphi\rangle\ ,
\end{equation}
and anti-commutation relations for Fermions,
\begin{equation}
\label{eq4}
\{a_\psi,a^\dag_\varphi\}\equiv a_\psi\,a^\dag_\varphi\,+\,a^\dag_\varphi\,a_\psi=\langle\psi\vert\varphi\rangle\ .
\end{equation}
Creation and annihilation operators are the building blocks for constructing firstly polynomials, involving products of powers of them, and eventually generic functions of them forming suitable algebras of operators.

Given two orthogonal subsets $\mathcal{H}_{1,2}$ of orthonormal modes $\{\vert\psi_j\rangle\}_{j=1}^{n_1}$ and $\{\vert\varphi_k\rangle\}_{k=1}^{n_2}$, we shall denote by 
$\mathcal{A}_{1,2}$ the algebras generated by the creation and annihilation operators $\Big\{a_{\psi_j},a^\dag_{\psi_j}\Big\}_{j=1}^{n_1}$, respectively  
$\Big\{a_{\varphi_k},a^\dag_{\varphi_k}\Big\}_{k=1}^{n_2}$. 
From the commutation relations, it follows that any pair of elements $A_1\in\mathcal{A}_1$, $A_2\in\mathcal{A}_2$ commute: $[A_1,A_2]=0$. 
Having abandoned the use of the tensor product structure inherent in the particle approach, one can now consistently  extend to systems of identical particles the characterization of entanglement through the lack of factorization of two point correlation functions for local operators as in~\eqref{eq2}. In this way, focusing on operator commutativity, one first generalizes
\begin{enumerate}
\item
the notion of locality for distinguishable particles to {\sl mode-locality} for identical particle systems by declaring local any product 
\begin{equation}
\label{modeloc}
A_{12}\,=\,A_1A_2\ ,\qquad A_{1}\in\mathcal{A}_{1}\ ,\  A_2\in\mathcal{A}_2\ ;
\end{equation}
\hskip - 1cm and then
\item
the notion of bipartite {\sl entanglement} for pure states, by declaring entangled  those state vectors $\vert\Psi\rangle$ for which 
\begin{equation}
\label{eq7}
\langle\Psi\vert A_1A_2\vert\Psi\rangle\,\neq\,\langle\Psi\vert A_1\vert\Psi\rangle\,\langle\Psi\vert A_2\vert\Psi\rangle\ ,
\end{equation}
for at least one couple $A_1$, $A_2$ of local operators.
\end{enumerate}
We shall refer to the pair of commuting algebras $\big(\mathcal{A}_1,\mathcal{A}_2\big)$ as to an {\sl algebraic bipartition}; certainly, both mode-locality and mode-entanglement 
depend on the choice of bipartition and the possible bipartitions are uncountably many, unlike in the 
standard distinguishable particle setting where $\mathcal{A}_1$ and $\mathcal{A}_2$ are essentially fixed to be the algebras 
$\mathcal{O}_1$ and $\mathcal{O}_2$ of the observables of particle $1$ and of 
particle $2$.
Furthermore, it turns out (see \cite{Benatti2020} and references therein) that a state vector is separable, that is non-entangled, with respect to a chosen algebraic bipartition $\big(\mathcal{A}_1,\mathcal{A}_2\big)$, if and only if
\begin{equation}
\label{eq8}
\vert\Psi_{\textnormal{sep}}\rangle=\mathfrak{f}^\dag_1\mathfrak{f}^\dag_2\,\vert \textnormal{vac}\rangle\ ,
\end{equation}
where $\mathfrak{f}^\dag_{1,2}$  are functions of the creation operators $\Big\{a^\dag_{\psi_j}\Big\}_{j=1}^{n_1}$ in $\mathcal{A}_1$, respectively 
$\Big\{a^\dag_{\varphi_k}\Big\}_{k=1}^{n_2}$ in $\mathcal{A}_2$. The operators  $\mathfrak{f}_{1,2}$ are chosen such that $\langle\textnormal{vac}\vert \mathfrak{f}_1\,\mathfrak{f}^\dag_1\vert\textnormal{vac}\rangle=\langle\textnormal{vac}\vert\,\mathfrak{f}_2\,\mathfrak{f}^\dag_2\vert\textnormal{vac}\rangle=1$ whence
\begin{eqnarray}
\nonumber
\langle\Psi_{\textnormal{sep}}\vert\Psi_{\textnormal{sep}}\rangle&=&\langle\textnormal{vac}\vert\mathfrak{f}_2\,\mathfrak{f}_1\,\mathfrak{f}^\dag_1\,\mathfrak{f}^\dag_2\vert\textnormal{vac}\rangle=
\langle\textnormal{vac}\vert \,\mathfrak{f}_1\,\mathfrak{f}^\dag_1\,\mathfrak{f}_2\,
\mathfrak{f}^\dag_2\vert\textnormal{vac}\rangle\\
\label{aita1}
&=&
\langle\textnormal{vac}\vert \mathfrak{f}_1\,\mathfrak{f}^\dag_1\vert\textnormal{vac}\rangle\,\langle\textnormal{vac}\vert\,\mathfrak{f}_2\,\mathfrak{f}^\dag_2\vert\textnormal{vac}\rangle=1\ .
\end{eqnarray}
\begin{rem}
\label{rem2}
Notice that, when the total number of Bosons is fixed, the number of particles in 
$\mathfrak{f}^\dag_{1,2}\vert\textnormal{vac}\rangle$ is also fixed. Hence,  the state vectors that are eigenstates of the total Boson number and are separable with respect to 
the bipartition $(\mathcal{A}_1,\mathcal{A}_2)$ must also be eigenstates of the number operators 
\begin{equation}
\label{numbops}
N_1=\sum_{j=1}^{n_1}a_{\psi_j}^\dag\,a_{\psi_j}^{\phantom{\dag}}\ ,\quad 
N_2=\sum_{k=1}^{n_2}a_{\varphi_k}^\dag\,a_{\varphi_k}^{\phantom{\dag}}
\end{equation} counting the Bosons in the modes that identify the commuting algebras $\mathcal{A}_{1,2}$.
\end{rem}

In the context of mode-entanglement  the two questions raised above  have a definite answer; indeed, in the second quantization formalism, the state in~\eqref{eq1} reads
\begin{equation}
\label{eq9}
\vert\Psi\rangle=a^\dag_{\chi}\,a^\dag_{\phi}\,\vert\textnormal{vac}\rangle\ ,
\end{equation}
and, according to~\eqref{eq8}, is thus separable with respect to an algebraic bipartition $\big(\mathcal{A}_1,\mathcal{A}_2\big)$ with $\mathfrak{f}^\dag_1=a^\dag_{\chi}$ and $\mathfrak{f}^\dag_2=a^\dag_{\phi}$ where $a_{\chi},a^\dag_{\chi}\in\mathcal{A}_1$ 
and $a_{\phi},a^\dag_{\phi}\in\mathcal{A}_2$.

Furthermore, once an orthonormal set $\{\vert\chi_j\rangle\}_j$ of modes in the single-particle Hilbert space
is given, the single particle observables 
$$
X=\sum_{i,j}X_{ij}\, \vert\chi_i\rangle\langle\chi_j\vert\ ,\quad Y=\sum_{k,\ell}Y_{k\ell}\, 
\vert\chi_k\rangle\langle\chi_\ell\vert\ ,
$$ 
in~\eqref{eq2}  have the following second quantized expressions (see~\eqref{eq2a})
$$
X_{\textnormal{sym}}=\sum_{i,j}X_{ij} \, a^\dag_{\chi_i}a_{\chi_j}\ ,\quad Y_{\textnormal{sym}}=\sum_{k,\ell}Y_{k\ell} \,a^\dag_{\chi_k}a_{\chi_\ell}\ .
$$
Then, if $X_{\textnormal{sym}}$ belongs to the algebra $\mathcal{A}_1$ corresponding to the set of modes $\{\vert\psi_j\rangle\}_{j=1}^{n_1}$ and $Y_{\textnormal{sym}}$ to the algebra $\mathcal{A}_2$ relative to the set of modes $\{\vert\varphi_k\rangle\}_{k=1}^{n_2}$, namely,
\begin{equation}
\label{eq9b}
X_{\textnormal{sym}}=\sum_{i,j=1}^{n_1}X_{ij} \, a^\dag_{\psi_i}a_{\psi_j}\ ,\quad X_{\textnormal{sym}}=\sum_{k,\ell=1}^{n_2}Y_{k\ell} \,a^\dag_{\varphi_k}a_{\varphi_\ell}\ ,
\end{equation}
the symmetrized observable $C$ in~\eqref{eq2} is local with respect to $\big(\mathcal{A}_1,\mathcal{A}_2\big)$; indeed, from the commutativity of $\mathcal{A}_1$ and $\mathcal{A}_2$  it follows that
\begin{align}
\nonumber
C & =\sum_{i,j=1}^{n_1}\sum_{k,\ell=1}^{n_2}\,\big(X\otimes Y\big)_{ik,j\ell}\, a^\dag_{\psi_i}\,a^\dag_{\varphi_k}\,a_{\varphi_\ell}\,a_{\psi_j}\\
\label{eq10}
& =
\left(\sum_{i,j=1}^{n_1}X_{ij}\, a^\dag_{\psi_i}\,a_{\psi_j}\right)\,\left(\sum_{k,\ell=1}^{n_2}Y_{k\ell}\,a^\dag_{\varphi_k}\,a_{\varphi_\ell}\right)\ .
\end{align}
It is also worth noting that, according to~\eqref{modeloc} and~\eqref{eq7},
mode-separable state vectors  
$\vert\Psi_{\textnormal{sep}}\rangle$  with respect to an algebraic bipartition $\big(\mathcal{A}_1,\mathcal{A}_2\big)$ as in~\eqref{eq8} remain separable under the action of mode-local operations transforming $\vert \Psi_{\textnormal{sep}}\rangle$ into 
\begin{equation}
\label{locops}
\vert\tilde \Psi\rangle:=O_1 O_2\vert \Psi_{\textnormal{sep}}\rangle\ ,\quad  O_{1,2}\in\mathcal{A}_{1,2}\ ,
\end{equation}
where $\vert\Psi_{\textnormal{sep}}\rangle$ is normalized as in \eqref{aita1}, while $O_1$ and $O_2$ are suitably rescaled 
to satisfy
$\langle \Psi_{sep}\vert O_1^\dag O_1\vert\Psi_{sep}\rangle=\langle \Psi_{sep}\vert O_2^\dag O_2\vert\Psi_{sep}\rangle=1$.
Indeed, the form~\eqref{eq8} of separable states is maintained by the action of $O_1O_2$; moreover, for arbitrary local observables $A_1A_2$, with $A_{1,2}\in\mathcal{A}_{1,2}$, setting $\tilde A_{1,2}=O_{1,2}^\dag A_{1,2}O_{1,2}$, from the fact that $O_1^\dag O_2^\dag A_1 A_2 O_2 O_1=O_1^\dag A_1 O_1\,O_2^\dag A_2 O_2$, 
using separability and the normalization conditions one derives
\begin{align}
\nonumber
&\langle\tilde\Psi\vert A_1A_2\vert\tilde\Psi\rangle
=\langle\Psi_{\textnormal{sep}}\vert\tilde A_1 \tilde A_2\vert\Psi_{\textnormal{sep}}\rangle=\langle\Psi_{\textnormal{sep}}\vert \tilde A_1\vert\Psi_{\textnormal{sep}}\rangle\,\langle\Psi_{\textnormal{sep}}\vert \tilde A_2\vert\Psi_{\textnormal{sep}}\rangle\\
\nonumber
&=\langle\Psi_{\textnormal{sep}}\vert \tilde A_1\vert\Psi_{\textnormal{sep}}\rangle\,
\langle\Psi_{\textnormal{sep}}\vert O_2^\dag O_2\vert\Psi_{\textnormal{sep}}\rangle\,\langle\Psi_{\textnormal{sep}}\vert \tilde A_2\vert\Psi_{\textnormal{sep}}\rangle\,\langle\Psi_{\textnormal{sep}}\vert O_1^\dag O_1\vert\Psi_{\textnormal{sep}}\rangle\\
\label{aita3}
&=\langle\Psi_{\textnormal{sep}}\vert \tilde A_1\,O_2^\dag O_2\vert\Psi_{\textnormal{sep}}\rangle\,
\langle\Psi_{\textnormal{sep}}\vert \tilde A_2\,O_1^\dag O_1\vert\Psi_{\textnormal{sep}}\rangle=\langle\tilde\Psi\vert A_1\vert\tilde\Psi\rangle\,\langle\tilde\Psi\vert A_2\vert\tilde\Psi\rangle\ .
\end{align}

\begin{rem}
\label{rem1}
The formulation of mode-entanglement in Fermionic systems  and the role of canonical anti-commutation relations is discussed in~\cite{Banuls2007,Benatti2014,Benatti2014-2,Benatti2016,Ding2020}.
In this case, the parity $(-1)^N$, where $N$ is the total number of Fermions, is a conserved
quantity, and so observable subalgebras can be decomposed in an even and an odd
part, which project $N$ onto the subspace with even and odd eigenvalues respectively.
The odd part of the subalgebras, $A_1$ and $A_2$, must anti-commute with each other
while the even part of each subalgebra must commute with both the even and the
odd part of the other subalgebra. The definitions of mode-local operators~\eqref{modeloc} and of mode-entangled pure states~\eqref{eq7}, as well as their general expression~\eqref{eq8}, remain unchanged: of course, Fermionic systems are constrained to have at most one particle
per mode.

Also, mode-entanglement can be extended to mixed states \cite{Benatti2020}, thereby generalizing Werner's reformulation of entanglement for distinguishable particles~\cite{Werner1989}, \textit{i.e.} equation \eqref{eq6}, where, as already stressed, algebras are made of single-particle operators. Mode-entanglement recovers Werner's definition when modes are occupied by at most one particle. Notice that other existing definitions of indistinguishable particle entanglement are not compatible with the Werner's formulation \cite{Benatti2020,Johann2020}. Indeed, for any choice of subalgebras, $\mathcal{A}_1$ and $\mathcal{A}_2$, there are operators $A_j\in\mathcal{A}_j$, $j=1,2$, and pure states termed separable within these approaches, that do exhibit correlations of the type embodied by inequality \eqref{eq7}.
\end{rem}

\subsection{Theoretical setting}

In the following we shall consider a simplified yet rather illustrative setting consisting of Bosons with spatial $S=L,R$
and internal, $\sigma=\uparrow,\downarrow$, degrees of freedom corresponding to the orthonormal basis 
\begin{equation}
\label{ONB}
\vert L,\uparrow\rangle\ ,\ \vert L,\downarrow\rangle\ ,\ \vert R,\uparrow\rangle\ ,\ \vert R,\downarrow\rangle\ ,
\end{equation}
in the single particle Hilbert space. 
One can then consider the corresponding mode creation and annihilation operators $a_{S,\sigma}$, $a^\dag_{S,\sigma}$ and proceed to bipartite the whole Bose algebra constructed upon such mode-operators into commuting algebras made of functions of the left or of the right mode-operators respectively, denoted, in a compact notation, by
\begin{equation}
\mathcal{A}_L=\{a_{L,\sigma},a_{L,\sigma}^\dag\}_{\sigma=\uparrow,\downarrow}\ ,\quad 
\mathcal{A}_R=\{a_{R,\sigma},a_{R,\sigma}^\dag\}_{\sigma=\uparrow,\downarrow}\ .
\end{equation}
Products $A_LA_R$ of observables $A_L\in\mathcal{A}_L$ and $A_R\in\mathcal{A}_R$ such as, for instance, the 
Boson number operators, 
\begin{equation}
\label{numbop}
N_L=\sum_{\sigma=\uparrow,\downarrow}a_{L,\sigma}^\dag a_{L,\sigma}\ ,\quad N_R=\sum_{\sigma=\uparrow,\downarrow}a_{R,\sigma}^\dag a_{R,\sigma}\ ,
\end{equation}
are mode-local with respect to the algebraic bipartition $\big(\mathcal{A}_L,\mathcal{A}_R\big)$. 

Other choices of commuting algebras, \textit{e.g.} $\mathcal{A}_\uparrow=\{a_{S,\uparrow},a_{S,\uparrow}^\dag\}_{S=L,R}$ and $\mathcal{A}_\downarrow=\{a_{S,\downarrow},a_{S,\downarrow}^\dag\}_{S=L,R}$, give rise to different kind of mode-entanglement.
In all cases, unlike particles which cannot in general be individually addressed, modes described by commuting algebras can.

Though reduced to few degrees of freedom, the above setting allows one to discuss the usefulness for quantum metrology and teleportation of $N$ Bosons occupying the states $\vert L,\uparrow\rangle$ and $\vert R,\downarrow\rangle$. 
These $N$-Boson states belong to a subspace spanned by state vectors that, in first quantization, read
\begin{equation} \label{symm}
\mathcal{S}\,\Big(\vert L,\uparrow\rangle^{\otimes\ell}\otimes\vert R,\downarrow\rangle^{\otimes(N-\ell)}\Big),
\end{equation}
where $\mathcal{S}$ is the symmetrization projector and, in second quantization, become Fock number states
\begin{equation}
\label{Psi1}
\vert \Psi_{\textnormal{Fock}}\rangle=\frac{\big(a_{L,\uparrow}^\dag\big)^\ell}{\sqrt{\ell!}}\,\frac{\big(a_{R,\downarrow}^\dag\big)^{N-\ell}}{\sqrt{(N-\ell)!}}\,|\textnormal{vac}\rangle,
\end{equation}
satisfying (see~\eqref{numbop})
\begin{equation}
\label{Psi1a}
N_L\vert \Psi_{\textnormal{Fock}}\rangle=\ell\,\vert\Psi_{\textnormal{Fock}}\rangle\ ,\quad N_R\,\vert  \Psi_{\textnormal{Fock}}\rangle=(N-\ell)\,\vert\Psi_{\textnormal{Fock}}\rangle\ .
\end{equation}
A generic normalized state vector spanned by the above Fock states reads
\begin{equation} 
\label{Psi}
\vert\Psi\rangle=\sum_{\ell=0}^N\alpha_\ell\,\frac{\big(a_{L,\uparrow}^\dag\big)^\ell}{\sqrt{\ell!}}\,\frac{\big(a_{R,\downarrow}^\dag\big)^{N-\ell}}{\sqrt{(N-\ell)!}}\,|\textnormal{vac}\rangle\ ,\qquad \alpha_\ell\in \mathbb{C}\ ,\quad \sum_{\ell=0}^N|\alpha_\ell|^2=1\ .
\end{equation}
Particular instances of these states can be manipulated and used as resources for quantum protocols; indeed, we propose to  focus upon the following states as they exemplify several resource states used in quantum information tasks:
\begin{enumerate}
\item
a superposition of two Fock number states,
\begin{eqnarray} 
\nonumber
|\Psi_{\ell_1,\ell_2}\rangle&=&\sqrt{\xi}\,\frac{\big(a_{L,\uparrow}^\dag\big)^{\ell_1}}{\sqrt{\ell_1!}}\,\frac{\big(a_{R,\downarrow}^\dag\big)^{N-\ell_1}}{\sqrt{(N-\ell_1)!}}\,|\textnormal{vac}\rangle\\
\label{Psi12}\hskip 2cm &+&e^{i\varphi}\sqrt{1-\xi}\,\frac{\big(a_{L,\uparrow}^\dag\big)^{\ell_2}}{\sqrt{\ell_2!}}\,\frac{\big(a_{R,\downarrow}^\dag\big)^{N-\ell_2}}{\sqrt{(N-\ell_2)!}}\,|\textnormal{vac}\rangle\ ,
\end{eqnarray}
with $0\leq\xi\leq 1$, that reduces to the so-called NOON state when $\ell_1=N-\ell_2=0$ and $\xi=1/2$;
\item
a linear superposition of Fock states with uniform weights $\displaystyle\frac{1}{N+1}$ and arbitrary phases, 
\begin{equation} \label{PsiUnif}
|\Psi_{\textnormal{unif}}\rangle=\sum_{\ell=0}^N\frac{e^{i\varphi_\ell}}{\sqrt{N+1}}\,\frac{\big(a_{L,\uparrow}^\dag\big)^\ell}{\sqrt{\ell!}}\,\frac{\big(a_{R,\downarrow}^\dag\big)^{N-\ell}}{\sqrt{(N-\ell)!}}\,|\textnormal{vac}\rangle\ ;
\end{equation}
\item
a coherent state of the SU(2) group generated by the two modes $(L,\uparrow)$ and $(R,\downarrow)$,
\begin{equation} \label{PsiCoh}
\vert\Psi_{\textnormal{coh}}\rangle=\frac{1}{\sqrt{N!}}\left(\sqrt{\xi}\,a_{L,\uparrow}^\dag+e^{i\varphi}\sqrt{1-\xi}\,a_{R,\downarrow}^\dag\right)^N|\textnormal{vac}\rangle\ ,\quad 0\leq\xi\leq 1\ .
\end{equation}
\end{enumerate}

With the exception of the Fock number state $\vert\Psi_{\textnormal{Fock}}\rangle$ which, according to~\eqref{eq8}, is mode-separable with respect to the bipartitions 
 $(\mathcal{A}_L,\mathcal{A}_R)$ and $\big(\mathcal{A}_\uparrow,\mathcal{A}_\downarrow\big)$, all other states are mode-entangled with respect to them.
Moreover, the state $\vert\Psi_{\textnormal{unif}}\rangle$ is maximally entangled as quantified by several entanglement measures, like entanglement entropy and entanglement negativity \cite{Benatti2012,Benatti2012-2}.

\section{Two quantum protocols} \label{protocols-sec}

In the following we study the resourcefulness of the three states introduced above, together with the Fock ones, with respect to two quantum tasks implementable in an identical particle context; the first task is a metrological one and the second a mode-teleportation one.
For both tasks, we shall study how mode-entanglement provides a quantum advantage in the first case and enables the implementation of the second one.

\subsection{Identical particle metrology}
 \label{metrology}

As a first protocol, let us consider the estimation of the phase-shift $\theta$ implemented by a Mach-Zender interferometer acting on an $N$-Boson state $\vert\Psi\rangle$ as in~\eqref{Psi}
by shifting the relative phase of the modes $\vert L,\uparrow\rangle$ and $\vert R,\downarrow\rangle$. 
The action of such an interferometer is described  by the unitary operator 
\begin{equation}
\label{unitinterf}
U_\theta={\rm e}^{i\theta\,N_{LR}} \ ,\quad N_{LR}:=N_{L,\uparrow}-N_{R,\downarrow}\ ,
\end{equation}
where $N_{L,\uparrow}=a_{L,\uparrow}^\dag a_{L,\uparrow}$ and $N_{R,\downarrow}=a_{R,\downarrow}^\dag a_{R,\downarrow}$ count the number of Bosons in the modes
$\vert L,\uparrow\rangle$ and $\vert R,\downarrow\rangle$.

In quantum metrology, the quantum Cram\'er-Rao inequality lower-bounds the best accuracy $\delta^2\theta$ reachable in the determination of $\theta$ by the inverse of the quantum Fisher information 
$F_{\vert \Psi\rangle}(N_{LR})$ relative to the input state $\vert\Psi\rangle$ and the generator $N_{LR}$ of the transformation~\cite{Helstrom1976,Holevo1982,Braunstein1996,Paris2009}. 
Being the input state pure, the quantum Fisher information reduces to (four times) the variance $\Delta_{\vert\Psi\rangle}N_{LR}$
of the generator with respect to the state $\vert\Psi\rangle$. Then
\begin{equation}
 \label{QKRB}
\delta^2\theta\geqslant\frac{1}{F_{|\Psi\rangle}(N_{LR})}=\frac{1}{4\,\Delta_{\vert\Psi\rangle}N_{LR}} \ ,\quad 
\Delta_{\vert\Psi\rangle}N_{LR}=\langle\Psi\vert N_{LR}^2\vert\Psi\rangle\,-\,\langle\Psi\vert N_{LR}\vert\Psi\rangle^2\ .
\end{equation}
The best accuracy achievable by classical interferometry is the inverse of the classical Fisher information; it corresponds to the so-called shot-noise, namely to a Fisher information increasing linearly with the number of particles.
Whenever the Fisher information exceeds the shot-noise threshold, quantum features, either in the initial state or in the interferometer, need be at work.

In the case of the metrological task depicted above, the required quantumness is the mode-entanglement of the state with respect to the 
bipartition $\big(\mathcal{A}_L,\mathcal{A}_R\big)$. 
In fact, according to~\eqref{modeloc}, the unitary operator $U_\theta$ in~\eqref{unitinterf} 
is mode-local with respect to the bipartition $\big(\mathcal{A}_L,\mathcal{A}_R\big)$, for $N_{L,\uparrow}$ and $N_{R,\downarrow}$ commute so that 
$U_\theta$ splits into a product of commuting unitary operators
\begin{equation}
\label{localunit}
U_\theta={\rm e}^{i\theta N_{L,\uparrow}}\,{\rm e}^{-i\theta N_{R,\downarrow}},\quad {\rm e}^{i\theta N_{L,\uparrow}}\in\mathcal{A}_L\ ,\quad {\rm e}^{-i\theta N_{R,\downarrow}}\in\mathcal{A}_R\ .
\end{equation}

It follows that mode-entanglement in the initial state is necessary to beat the shot-noise,  if the interferometer is mode-local and the particle number is conserved \cite{Tilma2010,Benatti2013}.
This fact can be best appreciated as follows: if the $N$-Boson state $\vert\Psi\rangle$ were mode-separable with respect to  
$\big(\mathcal{A}_L,\mathcal{A}_R\big)$ it would be of the form~\eqref{eq8} with $\mathfrak{f}^\dag_1=\mathfrak{f}^\dag_L$ a suitable function of $a^\dag_{L,\uparrow}$ and 
$a^\dag_{L,\downarrow}$ and $\mathfrak{f}^\dag_2=\mathfrak{f}^\dag_R$ a suitable function of $a^\dag_{R,\uparrow}$ and 
$a^\dag_{R,\downarrow}$.  Then, according to Remark~\ref{rem2} and using~\eqref{aita1}, one computes
\begin{equation}
\label{aita2a}
\langle\Psi\vert N^2_{LR}\vert\Psi\rangle=\langle \mathfrak{f}_L\vert N_{L\uparrow}^2\vert  \mathfrak{f}_L\rangle\,+\,\langle  \mathfrak{f}_R\vert N_{R\downarrow}^2\vert  \mathfrak{f}_R\rangle\,-\,
2\,\langle  \mathfrak{f}_L\vert N_{L\uparrow}\vert  \mathfrak{f}_L\rangle\,\langle  \mathfrak{f}_R\vert N_{R\downarrow}\vert  \mathfrak{f}_R\rangle\ ,
\end{equation}
where $\vert \mathfrak{f}_{L,R}\rangle\equiv\mathfrak{f}_{L,R}^\dag\vert\textnormal{vac}\rangle$. It thus follows that
\begin{equation}
\label{aita2}
F_{\vert \Psi\rangle}(N_{LR})=4\,\Delta_{\vert \Psi\rangle}N_{LR}=4\,\Delta_{\vert\mathfrak{f}_L\rangle}N_{L,\uparrow}+4\,\Delta_{\vert \mathfrak{f}_R\rangle}N_{R,\uparrow}= 0\ .
\end{equation}
Indeed, since the total number of Bosons is fixed to be $N$, $\vert  \mathfrak{f}_L\rangle$ and $\vert  \mathfrak{f}_R\rangle$ are as $|\Psi_\textnormal{Fock}\rangle$ in \eqref{Psi1}, hence 
eigenstates of the number operators $N_{L,\uparrow}$ and $N_{R,\downarrow}$. The vanishing variances are in agreement with the fact that phase-changes and particle number operators are conjugate observables so that perfect knowledge of one of them entails total ignorance about the other one.
This same situation occurs with any fixed, finite number of modes,
or for distinguishable particles with fixed and finite single-particle Hilbert space dimension.

For the mode-entangled states in~\eqref{Psi} one finds
\begin{equation} \label{imb}
N_{LR}\vert \Psi\rangle=\sum_{\ell=0}^N(2\ell-N)\,\alpha_\ell\,\frac{\big(a_{L,\uparrow}^\dag\big)^\ell}{\sqrt{\ell!}}\,\frac{\big(a_{R,\downarrow}^\dag\big)^{N-\ell}}{\sqrt{(N-\ell)!}}\,\vert\textnormal{vac}\rangle \ ,
\end{equation}
whence~\eqref{imb} togheter with~\eqref{QKRB} yields
\begin{equation}
F_{\vert \Psi\rangle}(N_{LR})=4\sum_{\ell=0}^N|\alpha_\ell|^2\,\ell^2-4\left(\sum_{\ell=0}^N|\alpha_\ell|^2\,\ell\right)^2\ .
\end{equation}
In the case of the specific states in~\eqref{Psi12}--~\eqref{PsiCoh} the Fisher information reads:
\begin{align}
\label{F12} F_{\vert\Psi_{\ell_1,\ell_2}\rangle}(N_{LR})= & 4\,\xi(1-\xi)(\ell_1-\ell_2)^2, \\
\label{Funif} F_{\vert\Psi_{\textnormal{unif}}\rangle}(N_{LR})= & \frac{1}{3}(N^2+2N)\\
\label{Fcoh} F_{\vert\Psi_{\textnormal{coh}}\rangle}(N_{LR})= & 4\,\xi(1-\xi)N\ .
\end{align}
It follows that $F_{\vert\Psi_{\ell_1,\ell_2}\rangle}(N_{LR})$ beats the shot-noise whenever $F_{\vert\Psi_{\ell_1,\ell_2}\rangle}(N_{LR})>N$, \textit{e.g.} 
for the NOON state when $\ell_1=N-\ell_2=0$ and $\xi=1/2$. In this case, $F_{\vert \Psi_{0,N}\rangle}(N_{LR})=N^2$ reaching the 
maximum value for the quantum Fisher information, the so-called Heisenberg limit. 
As for the other states, $F_{\vert\Psi_{\textnormal{unif}}\rangle}(N_{LR})$ always beats the shot-noise: indeed, if the phases $\varphi_\ell\propto\ell$, $\vert\Psi_{\textnormal{unif}}\rangle$ is the eigenstate of the phase operator canonically conjugated to $N_{LR}$ \cite{Vourdas1990}.
Instead, $F_{\vert\Psi_{\textnormal{coh}\rangle}}(N_{LR})$ never beats the shot-noise because 
$\vert\Psi_{\textnormal{coh}}\rangle$ is a coherent state of the SU(2) group whose elements model linear interferometers and reproduce classical performances. In particular, while equation~\eqref{aita2} shows that entanglement is necessary if the particle number is conserved \cite{Tilma2010,Benatti2013}, equation \eqref{Fcoh} implies that entanglement is not in general sufficient to beat the shot-noise with local interferometers. It is indeed well known also for distinguishable particles that not all entangled states are useful for quantum metrological advantages.

Now, let us consider the unitary operator 
\begin{equation}
\label{nonloc}
V_\theta:={\rm e}^{i\,\theta\,T_{LR}}\ ,\quad  T_{LR}:=a_{L,\uparrow}^\dag a_{R,\downarrow}+a_{R,\downarrow}^\dag a_{L,\uparrow}\ .
\end{equation}
It describes a transformation which is non-local with respect to the bipartition $\big(\mathcal{A}_L,\mathcal{A}_R\big)$ and amounts to a phase-change operated through a rotation of the modes $(L,\uparrow)$ and $(R,\downarrow)$,  implementable by means of beam splitters.
By acting  with $V_\theta$ on the state $\vert\Psi_{\textnormal{Fock}}\rangle$ in~\eqref{Psi1}, which we have seen to be mode-separable with respect to the bipartition  $\big(\mathcal{A}_L,\mathcal{A}_R\big)$, one can beat the shot-noise 
limit when $\ell\neq0,N$~\cite{Benatti2010,Benatti2011}. Indeed, one computes
\begin{equation}
F_{\vert\Psi_{\textnormal{Fock}}\rangle}(T_{LR})=4\,\Big(\langle\Psi_{\textnormal{Fock}}\vert T_{LR}^2\vert\Psi_{\textnormal{Fock}}\rangle-\langle\Psi_{\textnormal{Fock}}\vert T_{LR}\vert\Psi_{\textnormal{Fock}}\rangle^2\Big)=N+2\ell(N-\ell)\ ,
\end{equation}
which improves upon the shot-noise at least by a multiplicative constant for $\ell\neq0,N$, while reaches the Heisenberg scaling $F_{\vert \Psi_{\textnormal{Fock}}\rangle}(T_{LR})\sim N^2/2$ if $\ell\sim N/2$.
Instead, when $\ell=0,N$, the state $\vert \Psi_{\textnormal{Fock}}\rangle$ is of the coherent form of $\vert \Psi_{\textnormal{coh}}\rangle$ 
in~\eqref{PsiCoh} with $\xi=0,1$, that indeed ensures only classical performances. On the other hand, if $\ell\neq 0,N$, the resource able
to provide the entanglement in the output state necessary for quantum enhanced performances is the non-local action of the interferometer. It is therefore possible to achieve quantum advantages with mode-separable states of $N$ particles independently prepared in two non-overlapping localizations and internal states.
Namely, there is no need to entangle the input state if the devices operates non-locally on the modes being used.
Indeed, the theory of mode-entanglement identifies the entangling power of beam splitters as a quantum resource \cite{Gagatsos2013,Benatti2013,Hari-krishnan2020}, as also happens with violations Bell's inequalities in similar physical situations \cite{Yurke1992,Li2018}.

\subsection{Mode-teleportation protocol} 
\label{teleport}

As well known, quantum teleportation \cite{Bennett1993} is a protocol for the transmission of an unknown quantum state to a remote location using an entangled resource state, local operations and classical communications. In the standard framework of distinguishable particles, states of a specific particle, say particle $A$, are teleported to a distant one, say particle $C$. This latter particle is initially entangled with another particle, say particle $B$ that is close to particle $A$. The sender performs a projective measurements onto Bell states of the particles $A$ and $B$; then, according to the obtained result which is sent to him through a classical communication channel, the receiver applies a local operation to particle $C$, whose state finally turns out to coincide with the initial state of particle $A$. This protocol has been generalized to three indistinguishable particles by considering a fully symmetrized or antisymmetrized state~\cite{Marinatto2001}, whereby it was shown that symmetrization or antisymmetrization cannot consistently generate entanglement useful for implementing teleportation .

In many-body systems indistinguishable particles are not in general confined to different locations, otherwise they could be distinguished, neither can they be individually addressed, {\it e.g.} by ascribing them unambiguous properties.
A full generalization of the teleportation protocol in a many-body context consists in teleporting superpositions of occupation number eigenstates of suitable sets of modes in the Fock space, rather than particle states as done in a first quantization approach~\cite{Schuch2004-2,Heaney2009-2,Marzolino2015,Marzolino2016}. This general protocol recovers the standard teleportation protocols involving distinguishable particles when the local particle occupation numbers of these modes are fixed. 
Indeed, in this case, addressing modes coincides with addressing particles.

In order to be more specific, consider a state of $M$ indistinguishable Bosons distributed between the two modes $(X,\uparrow)$ and $(Y,\downarrow)$ with spatial locations $X$ and $Y$, according to the state
\begin{equation}
\label{Phi}
\vert\Phi\rangle=\sum_{\ell=0}^M c_\ell\,\frac{\big(a_{X,\uparrow}^\dag\big)^\ell}{\sqrt{\ell!}}\,\frac{\big(a_{Y,\downarrow}^\dag\big)^{M-\ell}}{\sqrt{(M-\ell)!}}\,|\textnormal{vac}\rangle\ ,
\end{equation}
where  the mode $(Y,\downarrow)$  plays the role of particle $A$ in the standard setting. The goal of the protocol presented below is to teleport particle states from 
the mode $(Y,\downarrow)$ to the mode $(R,\downarrow)$ which then corresponds to particle $C$ in the standard setting. Such a mode is accessible to $N$ bosons  described by a resource state $|\Psi\rangle$ as in equation \eqref{Psi} that is thus mode-entangled with respect to the algebraic bipartition $\big(\mathcal{A}_L,\mathcal{A}_R\big)$.
In this scheme the remaining mode $(L,\uparrow)$ plays thus the role of particle $B$ in standard teleportation.

The protocol consists in 
\begin{enumerate}
\item
a projective measurement onto Bell-like states with  fixed particle numbers relative to  the modes $(Y,\downarrow)$ and $(L,\uparrow)$ (see Appendix \ref{measurement}),
assuming that,  as much as particles $A$ and $B$ in the standard protocol, the location $L$ is close enough to the location $Y$  to allow for coherent manipulations of states; 
\item
a local operation on the mode $(R,\downarrow)$ conditioned to the Bell measurement outcome that needs to be communicated between the locations $L$ and $R$ using classical devices.
\end{enumerate}

The operations involved in the teleportation protocol are mode-local as they are described in terms of operators relative to the fixed spatial locations $Y$ and $L$ on the one hand, and $R$ on the other. The main difference with respect to the metrological setting  discussed in the previous Section is that the analysis of the teleportation protocol needs the extension of  the 
$2$-mode subalgebra, $\mathcal{A}_L$, to a $4$-mode one, $\mathcal{A}_{YL}=\{a_{Y,\sigma},a_{Y,\sigma}^\dag,a_{L,\sigma},a_{L,\sigma}^\dag\}_{\sigma=\uparrow,\downarrow}$. 

The teleportation performances are measured by the  \textit{fidelity}  given by the overlap between the input state 
$\vert\Phi\rangle$ and the obtained state of the modes $(X,\downarrow)$ and $(R,\uparrow)$ (as if they were in the same Hilbert space) averaged over all Bell measurement outcomes and over all $\vert\Phi\rangle$. For two-mode Boson states, the teleportation fidelity turns out to be \cite{Marzolino2015,Marzolino2016}
\begin{eqnarray}
\nonumber
&&
f(|\Psi\rangle)=\frac{2}{M+2}\left(1+\sum_{k\neq j,\,k,j=0}^N\frac{\textnormal{max}\{0,M+1-|k-j|\}}{2M+2}\,\times\right.\\
\label{fidelity}
&&\hskip 2.5cm
\left.\times\,\frac{\langle\textnormal{vac}|a_{R,\downarrow}^{N-k}a_{L,\uparrow}^k|\Psi\rangle\langle\Psi|\big(a_{L,\uparrow}^\dag\big)^k\big(a_{R,\downarrow}^\dag\big)^{N-j}|\textnormal{vac}\rangle}{\sqrt{k!(N-k)!j!(N-j)!}}\right).
\end{eqnarray}
If the resource state is $|\Psi_{\textnormal{unif}}\rangle$, the final state of modes $(X, \uparrow)$ and $(R, \downarrow)$ is the same as the state $|\Phi\rangle$ in~\eqref{Psi}  for a fraction $(N-M+1)/(N+1)$ of measurement outcomes, while other outcomes result in teleporting only some components of the same state. Thus, the teleportation fidelity approaches one for large particle number in the resource state:
\begin{equation}
f(|\Psi_{\textnormal{unif}}\rangle)=1-\frac{M}{3N+3}.
\end{equation}
Teleportation is not exact for all Bell measurement outcomes because of the particle number conservation. When the resource state is $\vert\Psi_{\textnormal{coh}}\rangle$ 
with $\xi=1/2$, the teleportation fidelity also approaches one when $N\to\infty$:
\begin{equation}
f(|\Psi_{\textnormal{coh}\rangle})=1-\mathcal{O}\left(\frac{M^2}{N}\right).
\end{equation}

It is remarkable that coherent states $|\Psi_{\textnormal{coh}}\rangle$ provide very good teleportation performances with negligible distortions for large $N$, although they behave classically for the estimation of interferometric phases.
Instead, the resource state $\vert\Psi_{\ell_1,\ell_2}\rangle$ gives rise to poor teleportation performances, very close to those of mode-separable states 
$\vert\Psi_{\textnormal{Fock}}\rangle$ in~\eqref{Psi1} that yields
\begin{equation}
f(|\Psi_{\textnormal{Fock}\rangle})=\frac{2}{M+2}.
\end{equation}
This fidelity equals the maximum fidelity for teleporting $(M+1)$-level distinguishable particles; further, note that also the mode $(Y,\downarrow)$ to be teleported by means of the protocol discussed here has $M+1$ orthogonal states.
The case of $N=M=1$ is particularly suited to enlighten how mode-entanglement is able to capture the role of indistinguishability. Indeed, in such a case, the states $|\Psi\rangle$ and $|\Phi\rangle$ are single-particle states and the Bell-like measurement outcomes that provide perfect teleportation are associated to projections onto single-particle states. It is therefore clear that correlations with respect to the mode-algebraic bi-partitions, rather than among particles, is the relevant resource (see Appendix \ref{app:telep1}).

The above protocol can also be applied to swap entanglement: the modes $(X,\uparrow)$ and $(R,\downarrow)$ in the initial state are not entangled, while they are after the protocol. If the aim is to distribute as much entanglement as possible, a measure of the entangled shared by the modes $(X,\uparrow)$ and $(R,\downarrow)$ is a better figure of merit than the teleportation fidelity as the latter carries information that is not relevant for entanglement distribution \cite{Marzolino2015,Marzolino2016}.

\section{Discussion} \label{disc-sec}

We have shown that, in two specific quantum metrological and teleportation 
tasks based on systems of indistinguishable particles, mode-entanglement is the quantum resource that allows for enhanced performances. Indeed, mode-entanglement is the natural generalization to identical particles of the notion of entanglement  of distinguishable particles; it provides a useful resource to beat the limitations due to local operations and classical communications. 
According to the discussion at the end of Section~\ref{mode-ent-sec},
mode-separable state vectors  with respect to an algebraic bipartition remain separable under the action of  operations that are mode-local with respect to the same bipartition, while classical communications can only generate statistical mixtures of states resulting from local operations \cite{Chitambar2014}.

The notion of locality that appears very naturally in a second quantization context cannot be consistently formulated
in other approaches to indistinguishable particle entanglement \cite{Benatti2020,Johann2020}. It is therefore interesting to compare the performances of the above quantum information tasks within other entanglement theories.
In particular, we consider two different approaches to identical particle entanglement that are based not on the notion of modes, rather on that of particles.

\subsection{No-label approach}

The no-label approach characterizes the entanglement of $N$ indistinguishable particles with at least two degrees of freedom within a formalism that avoids the use of unphysical particle labels \cite{LoFranco2016,Compagno2018}; as a consequence state vectors of particles with only one degree of freedom turn out to be always unentangled. We will adopt a reformulation in second quantization that is more convenient especially for many-particle states (see \cite{Compagno2018,Benatti2020}).
In this approach, the entanglement shared between $n$ particles and the remaining $N-n$ ones in an $N$-Boson  state vector  $\vert\Psi\rangle$ is  $i)$ defined  with respect to a given choice of a finite-dimensional subspace $\mathcal{K}$ of the single-particle Hilbert space, and $ii)$ evaluated by means of the von Neumann entropy of a matrix $\rho^{(n)}$ that is derived from the $N$-particle projector $\vert\Psi\rangle\langle\Psi\vert$ as follows. An orthonormal basis 
$\{\vert\psi_j\rangle\}_{j=1}^p$ is selected in the 
subspace $\mathcal{K}$; then, a selective measurement is performed based on the set of operators $\prod_{j=1}^p a_{\psi_j}^{n_j}$ with all possible choices of 
occupation numbers of the corresponding modes such that $\sum_{j=1}^p\,n_j=n$, transforming $\vert\Psi\rangle\langle\Psi\vert$ into
\begin{equation} 
\label{rho.n}
\rho^{(n)}=\frac{\displaystyle\sum_{n_1+\cdots n_p=n}
\left(\prod_{j=1}^p a^{n_j}_{\psi_j}\right)\vert\Psi\rangle\langle\Psi\vert\left(\prod_{k=1}^p \big(a^{n_k}_{\psi_k}\big)^\dag\right)}
{{\displaystyle\sum_{n_1+\cdot+n_p=n}\langle\Psi\vert\prod_{j=1}^n \big(a^{n_j}_{\psi_j}\big)^\dag a_{\psi_j}^{n_j}\vert\Psi\rangle}} \ .
\end{equation}
As regards $\rho^{(n)}$ two facts need be noticed: 
\begin{enumerate}
\item
while the von Neumann entropy of $\rho^{(n)}$ does depend on the chosen subspace $\mathcal{K}$, it does not depend on the chosen orthonormal basis of $\mathcal{K}$;
\item
the operators $\prod_{j=1}^p a^{n_j}_{\psi_j}$ remove $n$ particles from the $N$-particle state 
$\vert\Psi\rangle$. Therefore, $\rho^{(n)}$ is supported by the Fock sector with $N-n$ particles; however it is not an $N-n$ particle reduced 
density matrix as would result from a partial trace with respect to the discarded degrees of freedom: for instance, it does not reproduce expectations of $(N-n)$-particle observables \cite{Benatti2017}.
\end{enumerate}

In order to make a comparison between the mode and the no-label approaches to identical particle entanglement in the case of the metrological task, let $\mathcal{K}$ be the two-dimensional subspace of $\mathbb{C}^4$ spanned by the vectors $\vert L,\downarrow\rangle$ and $\vert L,\uparrow\rangle$. This latter is the usual choice in the applications of the no-label approach, which aims at addressing quantum correlations between localized particles.  
The matrix $\rho^{(n)}$ results then
\begin{equation}
\rho^{(n)}=\frac{\displaystyle\sum_{k=0}^n a^k_{L,\uparrow}\,a^{n-k}_{L,\downarrow}\,\vert\Psi\rangle
\langle\Psi\vert\,\big(a^{n-k}_{L,\downarrow}\big)^\dag\,\big(a^k_{L,\uparrow}\big)^\dag}
{\displaystyle\sum_{k=0}^n\langle\Psi\vert\big(a^k_{L,\uparrow}\big)^\dag a^k_{L,\uparrow}\,\big(a^{n-k}_{L,\downarrow}\big)^\dag a^{n-k}_{L,\downarrow}\vert\Psi\rangle}\ .
\end{equation}
In the case of states $\vert\Psi\rangle$ of the form~\eqref{Psi}, one finds
\begin{equation}
\label{aita4}
a^k_{L,\uparrow}\,a^{n-k}_{L,\downarrow}\,\vert\Psi\rangle=\delta_{kn}\sum_{\ell=n}^N\alpha_\ell \, \frac{a_{L,\uparrow}^n\,\big(a^\dag_{L,\uparrow}\big)^\ell}{\sqrt{\ell!}}
\frac{\big(a^\dag_{R,\downarrow}\big)^{N-\ell}}{\sqrt{(N-\ell)!}}\,\vert\textnormal{vac}\rangle\ ,
\end{equation}
whence
\begin{equation}
\label{aita5}
\rho^{(n)}=\vert\Psi^{(n)}\rangle\langle\Psi^{(n)}\vert\ ,\ \vert\Psi^{(n)}\rangle={\frac{1}{\sqrt{\displaystyle \sum_{\ell=n}^{N}\vert\alpha_\ell\vert^2}}}\,\sum_{\ell=n}^{N}
\alpha_\ell\,\frac{\big(a_{L,\uparrow}^\dag\big)^{\ell-n}}{\sqrt{(\ell-n)!}}\,\frac{\big(a_{R,\downarrow}^\dag\big)^{N-\ell}}{\sqrt{(N-\ell)!}}\,|\textnormal{vac}\rangle \ .
\end{equation}
In this case, the matrix $\rho^{(n)}$ is thus a pure state and therefore its von Neumann entropy is zero for any $n$. As a consequence, according to the no-label approach, the states $\vert\Psi\rangle$ are not entangled with respect to any particle partitioning.

Within the no-label approach, as much as the notion of entanglement, also that of locality depends on the single-particle subspace $\mathcal{K}$.
In particular, single-particle operators whose measurement projects single-particle states onto $\mathcal{K}$ are deemed local \cite{LoFranco2016,Castellini2019}. 
Then, it follows that the unitary operator $U_\theta$ in~\eqref{unitinterf} that implements the action of the
interferometer discussed in the metrological context is local being the product of the two spatially local phase shifts in~\eqref{localunit}. The puzzle here is thus as follows: 
in the mode-entanglement approach, the metrological quantum advantage is due to the entanglement of the state relative to the algebraic bipartition with respect to which the interferometer action is instead local. On the contrary, in the no-label approach, neither the state is entangled nor the interferometer is non-local, 
yet the quantum advantage is there as predicted 
by the behaviour of the Fisher information which is only due to the structure of the state $\vert\Psi\rangle$ and of the unitary operator $U_\theta$.
An experimental evidence that the shot-noise limit is beaten would then force one  to accept that, in the no-label approach, quantum metrological advantages might occur without 
either quantum entanglement or quantum non-locality during any step of the protocol.

Analogously, in the no-label approach, operators of the form
\begin{equation}
\label{bellop1}
\mathbbm{1}\otimes O_L\,+\,O_L\otimes \mathbbm{1}\quad \hbox{with}\quad O_L:= \vert L\rangle \langle L\vert \, O
\end{equation}
and $O$ a single particle operator, in first quantization, or 
\begin{equation}
\label{bellop2}
\sum_{\sigma,\sigma'=\uparrow,\downarrow}\langle\sigma|O|\sigma'\rangle\,a_{L,\sigma}^\dag\,a_{L,\sigma'}\ ,
\end{equation}
in second quantization, are local. Such structures can be straightforwardly generalized so that $n$-particle local operators have the form
\begin{equation}
\sum_{\sigma_j,\sigma'_j=\uparrow,\downarrow}\langle\sigma_1|\otimes\langle\sigma_2|\dots\otimes\langle\sigma_n|O|\sigma'_1\rangle\otimes|\sigma_2'\rangle\dots\otimes|\sigma_n'\rangle\prod_{j=1}^n a_{L,\sigma_j}^\dag a_{L,\sigma'_j}.
\end{equation}
Therefore, the measurements  in the teleportation protocol involve local observables in the $L$ and $Y$ positions, and thus act locally on the states $\vert\Psi\rangle$.
Furthermore, also the last step of the teleportation protocol is local as it acts unitarily with respect to the mode $(R,\downarrow)$ and possibly on nearby auxiliary modes. 
Like in the case of phase estimation with local interferometers, the teleportation protocol interpreted within the no-label approach is apparently implementable without either entanglement or quantum non-locality, quite a paradoxical conclusion.

\subsection{Particle-entanglement: another approach}

Another particle-based approach to entanglement in systems of indistinguishable particles, sometimes called particle-entanglement, defines as separable states those where all particles are in the same single-particle state $\vert\psi\rangle$, namely $\vert\psi\rangle^{\otimes N}$  in first quantization, or $\displaystyle\frac{(a^\dag)^N\vert\textnormal{vac}\rangle}{\sqrt{N!}}$ for any mode-creation operator $a^\dag$ in second quantization~ \cite{Paskauskas2001,Eckert2002,Grabowski2011-2,Morris2020}.

This approach is a straight extension of the standard theory of entanglement to identical particles. Indeed, unlike  in Section~\eqref{mode-ent-sec}, symmetrization is taken as a legitimate source of entanglement. For instance, in this way all Fermionic state vectors turn out to be entangled.
Within this approach, we  show that quantum advantages as teleportation emerge without
particle-entanglement, for they are obtainable from particle-separable states. It then follows that this particular type of particle-entanglement 
is not consistent with the operational point of view that asks for entanglement in order to achieve quantum advantages.

As a first example, let us consider  the teleportation protocol discussed in Section~\ref{teleport} and set $N=M=1$. Then the resource state~\eqref{Psi} reduces to $\vert\Psi\rangle=\big(a_{L,\uparrow}^\dag+a_{R,\downarrow}^\dag\big)|\textnormal{vac}\rangle$, and thus 
to the states $\vert\Psi_{\ell_1=0,\ell_2=1}\rangle$, $\vert\Psi_{\textnormal{unif}}\rangle$ 
and $\vert\Psi_{\textnormal{coh}}\rangle$ in~\eqref{Psi12},~\eqref{PsiUnif} and~\eqref{PsiCoh} with $N=1$. The state $\vert\Psi\rangle$ is  mode-entangled, but 
as a single-particle state it is trivially  not particle-entangled.
On the other hand, the protocol uses projectors onto single-particle states that provide probabilistically  perfect teleportation, but, as such, cannot generate particle-entanglement.
Therefore, the particle-entanglement so far considered  is not necessary for teleportation.

As a second instance where paradoxical results emerge adopting the above definition of particle-entanglement, let us consider a slightly different teleportation protocol which aims at distributing entanglement at distant locations $X$ and $R$, as done in entanglement swapping.
The protocol hinges upon an input state of coherent form,
\begin{equation}
\label{phi2}
\vert\Phi_{\textnormal{coh}}\rangle=\frac{1}{\sqrt{M}}\left(\sqrt{\zeta}\,a_{X,\uparrow}^\dag+e^{i\vartheta}\sqrt{1-\zeta}\,a_{Y,\downarrow}^\dag\right)^M|\textnormal{vac}\rangle,
\end{equation}
and upon the resource state $|\Psi_{\textnormal{coh}}\rangle$ as in~\eqref{PsiCoh}. Both these states are particle-separable according to the above definition of particle-entaglement. 
Let us take as protocol global initial state 
the combination
$\vert\Phi_{\textnormal{coh}}\rangle|\Psi_{\textnormal{coh}}\rangle$; notice that this state
is not particle-separable since it is not the product of same single-particle states, despite the fact that
$|\Phi_{\textnormal{coh}}\rangle$ and $|\Psi_{\textnormal{coh}}\rangle$ can be prepared independently at distant locations without any interactions between the corresponding particles.
In addition, let us replace the Bell-like measurement with the following generalized measurement:
\begin{equation}
\rho\longrightarrow\frac{E_i\rho E_i^\dag}{\textnormal{Tr}(\rho E_i^\dag E_i)}\ ,
\end{equation}
with outcomes $i\in\{0,1\}$ and
\begin{align}
\label{E0} E_0 & =|\eta,\omega\rangle\langle\eta,\omega|\ , \qquad |\eta,\omega\rangle=\frac{1}{\sqrt{n}}\left(\sqrt{\eta}\,a_{Y,\downarrow}^\dag+e^{i\omega}\sqrt{1-\eta}\,a_{L,\uparrow}^\dag\right)^n|\textnormal{vac}\rangle\ , \\
E_1 & =\mathbbm{1}-E_0\ ,
\end{align}
where $0\leq\eta\leq 1$ and $\omega$ an arbitrary phase. Note that only $E_1$ can generate entanglement as 
$E_0$ projects onto a particle-separable state.
Nevertheless, by discarding the ``1'' outcome and concentrating on the ``0'' one, as outcome of the protocol one ends up with the state:
\begin{equation}
\frac{E_0|\Phi_{\textnormal{coh}}\rangle|\Psi_{\textnormal{coh}}\rangle}{\sqrt{\langle\Phi_{\textnormal{coh}}|\langle\Psi_{\textnormal{coh}}|E_0^\dag E_0|\Phi_{\textnormal{coh}}\rangle|\Psi_{\textnormal{coh}}\rangle}}\ ,
\end{equation}
with probability
\begin{equation}
\langle\Phi_{\textnormal{coh}}|\langle\Psi_{\textnormal{coh}}|E_0^\dag E_0|\Phi_{\textnormal{coh}}\rangle|\Psi_{\textnormal{coh}}\rangle\ .
\end{equation}
For $N=M=n$, one explicitly finds:
\begin{align}
& E_0|\Phi_{\textnormal{coh}}\rangle|\Psi_{\textnormal{coh}}\rangle= \nonumber \\
& =|\eta,\omega\rangle\frac{1}{\sqrt{N!}}\sum_{k=0}^N{N \choose k}^2\big(\eta(1-\xi)(1-\zeta)e^{i(\vartheta+\varphi)}\big)^{\frac{N-k}{2}}\times\\
&\hskip 1cm \times
\big(\xi\zeta(1-\eta)e^{-i\omega}\big)^{\frac{k}{2}}
\big(a_{X,\uparrow}^\dag\big)^k
\big(a_{R,\downarrow}^\dag\big)^{N-k}|\textnormal{vac}\rangle.
\end{align}
This final state turns out to be in general entangled with respect to the untouched modes $(X,\uparrow)$ and $(R,\downarrow)$. Therefore, the above swapping protocol probabilistically generates entanglement 
starting from a state, $|\Phi_{\textnormal{coh}}\rangle |\Psi_{\textnormal{coh}}\rangle$, that, as already stressed, can be prepared using only local operations, quite a striking paradox.
Such an inconsistency originates from the impossibility of reconciling the above definition of particle-entanglement with any underlying notion of identical particle locality (see Remarks~\ref{loc:rem} and~\ref{rem1}).

A resource theory associated with the notion of particle-entanglement  
here discussed has been recently proposed \cite{Morris2020}.
In general,
resource theories describe quantum resources beyond entanglement \cite{Chitambar2019}; they are based on defining resourceless states, called \textit{free states}, and resourceless operations, called \textit{free operators}, as a subset (sometimes a proper subset as in the case of resource theories of entanglement \cite{Chitambar2014},\cite{Chitambar2020} and quantum coherence \cite{Streltsov2017}) of operators that cannot transform free states into resourceful ones.
It turns out that the resource theory proposed in \cite{Morris2020} identifies states of the form
$|\Phi_{\textnormal{coh}}\rangle |\Psi_{\textnormal{coh}}\rangle$ as resourceful, but without specifying any notion 
of particle-local operations. Instead, in discussing entanglement theory, the notion of locality is crucial in the definition of free operations, that typically consist of local operations and classical communication \cite{Chitambar2014}.

\section{Conclusions} \label{conclusions}

We have discussed a fully consistent generalization of entanglement theory to systems of indistinguishable particles and some of  its applications to quantum information processing. Such a general theory, namely mode entanglement, is formulated in second quantization and accounts for quantum correlations between modes in the Fock space. Indeed, identical particles are not individually addressable, unless they are effectively distinguished by unanbiguous properties, {\it i.e.} orthogonal states of certain degrees of freedom. On the other hand, modes are experimentally addressable \cite{Wurtz2009,Bakr2010,Sherson2010}, and mode addressability retrieves particle addressability when identical particles become effectively distinguishable~\cite{Benatti2020}. In general, however, full physical consistency requires that the relevant degrees of freedom in many-body quantum information processing be modes rather than particles.

We have shown that mode entanglement is the relevant quantum resource for interferometric phase estimation and teleportation, two benchmark protocols that are central in quantum technologies. Indeed, indistinguishability improves the capability of these protocols with respect to those making use of distinguishable particles. First of all, identical particles can be distinguished at the cost of sacrificing certain degrees of freedom that can only be used as particle labels and not as an active part of the information processing. Secondly, indistinguishable particles lift off the load to generate entanglement at the input of an interferometer in order to estimate the phase with quantum enhanced accuracy, the reason being that indistinguishability allows the interferometer itself to generate mode entanglement. Finally, mode entanglement in systems of indistinguishable particles enables the implementation of mode teleportation that thus generalizes standard teleportation.

We have also compared the above protocols with alternative approaches to indistinguishable particle entanglement within the context of resource analysis, pointing out logical contradictions that arise in adopting those definitions. In particular, the no-label approach that introduces an {\it ad hoc} formalism in order to avoid unphysical particle labels, attributes no role to entanglement in the realization of the protocols described above. Instead, the other discussed approach to particle entanglement attributes entangling power to the symmetrization of identical particle pure states. As such, it entails that teleportation can be used to generate entanglement between distant particles using only non-entangling operations.

These results clearly pinpoint internal inconsistencies in those definitions of ``particle entanglement'',
making them unreliable in their physical predictions. Instead, in all so-far tested physical applications, no inconsistencies emerge
using second-quantized, mode entanglement in describing quantum correlations in system of identical particles: indeed, this apprach is able to correctly identify entanglement as a crucial ingredient in the realization of quantum protocols.

We hope that these findings will stimulate further research on the use of many-body systems consisting of identical constituents in quantum information processing, and in particular encourage the development of
quantum technological advances able to experimentally confirm
the above discussed results.

\appendix

\section{Bell-like measurement} \label{measurement}

The Bell-like measurement in the teleportation protocol is a generalization of that used in the teleportation of qudits \cite{Bennett1993}. It consists of projectors onto the following states
\begin{equation} 
\label{B}
\vert B_{\ell,\lambda}\rangle_{YL}=\sum_{k=0}^M\frac{e^{2\pi i\frac{\lambda k}{M+1}}}{\sqrt{M+1}}\,\frac{\big(a_{Y,\downarrow}^\dag\big)^{M-k}}{\sqrt{(M-k)!}}\cdot\frac{\big(a_{L,\uparrow}^\dag\big)^{k+\ell}}{\sqrt{(k+\ell)!}}\,\vert\textnormal{vac}\rangle\ , 
\end{equation}
where $\ell\in\{0,1,\cdots,N-M\}$ and $\lambda\in\{0,1,\cdots,M\}$.

The formal difference with respect to the Bell measurement for qudit teleportation can be appreciated by re-writing 
$$
\big(a_{Y,\downarrow}^\dag\big)^{M-k}\big(a_{L,\uparrow}^\dag\big)^{k+l}|\textnormal{vac}\rangle=\sqrt{(M-k)!(k+l)!}\,|M-k,k+l\rangle_{YL}\ .
$$
Indeed, the sum $k+\ell$ is taken modulo $N$ for qudits and the dimensions of each qudit Hilbert space are the same, while, in the Fock space, the qudit measurement violates the conservation of the particle number that is imposed here. In order to overcome this difficulty, we consider the standard sum $k+\ell$ 
and allow for larger occupation numbers of the mode $(L,\uparrow)$. Despite this generalization, the states \eqref{B} do not span the state space available for the modes $(Y,\downarrow)$ and $(L,\uparrow)$, and the corresponding measurement is not complete. Nevertheless, the ratio between the space spanned by \eqref{B} and the required space of the modes $(Y,\downarrow)$ and $(L,\uparrow)$ is $\displaystyle\frac{N-M+1}{N+1}$ approaching one for $N\gg M$.
Also, measurements with outcomes $(\ell,\lambda)$ projecting onto states \eqref{B} provide perfect teleportation when using the resource state $|\Psi_{\textnormal{unif}}\rangle$.

A complete measurement including projectors onto \eqref{B} is a projective measurements onto the following states
\begin{align} 
& \vert B_{\ell,\lambda}\rangle_{YL}=\sum_{k=\textnormal{max}\{0,-\ell\}}^{\textnormal{min}\{M,N-\ell\}}\frac{e^{2\pi i\frac{\lambda k}{\mathcal{C}_\ell}}}{\sqrt{\mathcal{C}_\ell}}\,\frac{\big(a_{Y,\downarrow}^\dag\big)^{M-k}}{\sqrt{(M-k)!}}\cdot\frac{\big(a_{L,\uparrow}^\dag\big)^{k+\ell}}{\sqrt{(k+\ell)!}}\,\vert\textnormal{vac}\rangle, \nonumber \\
\nonumber
&\\
& \mathcal{C}_\ell=
\begin{cases}
M+\ell+1 & \textnormal{if } -M\leqslant \ell<0 \\
M+1 & \textnormal{if } 0\leqslant l\leqslant N-M \\
N-\ell+1 & \textnormal{if } N-M<\ell\leqslant N \\
\end{cases}, \nonumber \\
\nonumber
&\\
\label{CB}
& \ell\in\{-M,-M+1,\cdots,N\}, \qquad \lambda\in\{0,1,\cdots,\mathcal{C}_\ell-1\}.
\end{align}
Projectors onto states \eqref{CB} not included in \eqref{B} result in the teleportation only of some components of input states $\vert\Psi\rangle$, and slightly increase the teleportation fidelity.

\section{Teleportation for distinguishable particle} \label{app:telep1}

For a better comparison with the case of distinguishable particle, it is instructive to set $N=M=1$ and rephrase the teleportation protocol presented in the previous Appendix within the first quantization approach. The input and the resource states are, respectively,
\begin{equation}
\vert\Phi\rangle=c_0|X,\uparrow\rangle+c_1|Y,\downarrow\rangle\in\mathcal{H}_{\textnormal{input}}, \qquad \vert \Psi\rangle=\alpha_0|R,\downarrow\rangle+\alpha_1
| L,\uparrow\rangle\in\mathcal{H}_{\textnormal{resource}},
\end{equation}
and the Bell-like measurement outcomes that provide perfect teleportation correspond to projections  onto the states
\begin{equation}
|B_{\pm}\rangle=\frac{|Y,\downarrow\rangle\pm|L,\uparrow\rangle}{\sqrt{2}}.
\end{equation}
Whether these projectors act either on the input state or on the resource state, it is not prossible to reproduce the desired output state, namely
\begin{equation}
c_0\vert X,\uparrow\rangle+c_1\vert R,\downarrow\rangle.
\end{equation}
Indeed, standard teleportation protocols for distinguishable particle states requires two- (or many-) particle resource states and Bell measurements. Teleportation in the framework of mode-entanglement recovers the standard protocol for fixed particle number in locations $Y$, $L$, and $R$ \cite{Marzolino2015,Marzolino2016,Benatti2020}. An example is provided by the following input state and resource state
\begin{eqnarray}
\vert\Phi\rangle&=&c_0\vert Y,\uparrow\rangle+c_1\vert Y,\downarrow\rangle\in\mathcal{H}_{\textnormal{input}}\ ,\\ 
\vert\Psi\rangle&=&\frac{\vert L,\downarrow\rangle\otimes|R,\downarrow\rangle+\vert L,\uparrow\rangle\otimes\vert R,\uparrow\rangle}{\sqrt{2}}\in\mathcal{H}_{\textnormal{resource}}\ ,
\end{eqnarray}
and by projectors onto the following states
\begin{eqnarray}
\vert B_{\pm}\rangle&=&\frac{\vert Y,\downarrow\rangle\otimes\vert L,\downarrow\rangle\pm|Y,\uparrow\rangle\otimes|L,\uparrow\rangle}{\sqrt{2}}\ ,\\
\vert B_{\pm}'\rangle&=&\frac{|Y,\downarrow\rangle\otimes\vert L,\uparrow\rangle\pm\vert Y,\downarrow\rangle\otimes\vert L,\uparrow\rangle}{\sqrt{2}}.
\end{eqnarray}

\section*{Acknowledgements}
U.~M. is financially supported by the European Union's Horizon 2020 research and innovation programme under the Marie Sk\l odowska-Curie grant agreement No. 754496 - FELLINI.
F. B., R. F. and U. M.
acknowledge that their research has been conducted within the framework of the Trieste Institute for Theoretical Quantum Technologies.



\end{document}